\documentclass[graybox]{svmult}

\usepackage{type1cm}        % activate if the above 3 fonts are
\usepackage{makeidx}         % allows index generation
\usepackage{graphicx}        % standard LaTeX graphics tool
\usepackage{multicol}        % used for the two-column index
\usepackage[bottom]{footmisc}% places footnotes at page bottom
\usepackage{natbib}          % additional package loaded by Michael

\usepackage{newtxtext}       % 
\usepackage{newtxmath}       % selects Times Roman as basic font
\usepackage{mathtools}         % additional package loaded by Michael
\usepackage{amsmath}

\makeindex             % used for the subject index
\def\ba{{\bf a}}

\def\br{{\bf r}}
\def\bc{{\bf c}}
\def\be{{\bf e}}

\def\bx{{\bf x}}
\def\by{{\bf y}}
\def\bone{{\bf 1}}

\def\bP{{\bf P}}
\def\bQ{{\bf Q}}

\def\bL{{\bf L}}

\def\bD{{\bf D}}

\def\bZ{{\bf Z}}

\def\bZ{{\bf Z}}
\def\bX{{\bf X}}
\def\bY{{\bf Y}}
\def\bI{{\bf I}}

\def\bGam{{\mathbf\Gamma}}

\def \tr{^{\sf\scriptsize \raisebox{0.3ex}{\sf T}}\mkern-1mu}

\begin{document}

\title*{Compositional data analysis --- linear algebra, visualization and interpretation}
\author{Michael Greenacre}
\institute{Michael Greenacre \at Universitat Pompeu Fabra, \email{michael.greenacre@upf.edu}}
\maketitle

\abstract*{Compositional data analysis is concerned with multivariate data that have a constant sum, usually 1 or 100\%.
These are data often found in biochemistry and geochemistry, but also in the social sciences, when relative values are of interest rather than the raw values.
Recent applications are in the area of very high-dimensional ``omics'' data.
Logratios are frequently used for this type of data, i.e. the logarithms of ratios of the components of the data vectors.
These ratios raise interesting issues in matrix-vector representation, computation and interpretation, which will be dealt with in this chapter.}

\abstract{Compositional data analysis is concerned with multivariate data that have a constant sum, usually 1 or 100\%.
These are data often found in biochemistry and geochemistry, but also in the social sciences, when relative values are of interest rather than the raw values.
Recent applications are in the area of very high-dimensional ``omics'' data.
Logratios are frequently used for this type of data, i.e. the logarithms of ratios of the components of the data vectors.
These ratios raise interesting issues in matrix-vector representation, computation and interpretation, which will be dealt with in this chapter.}

\section{Introduction}
\label{sec:1}
\begin{figure}[h!]
%\sidecaption[t]
\hspace{-0.2cm}
\includegraphics[scale=.54]{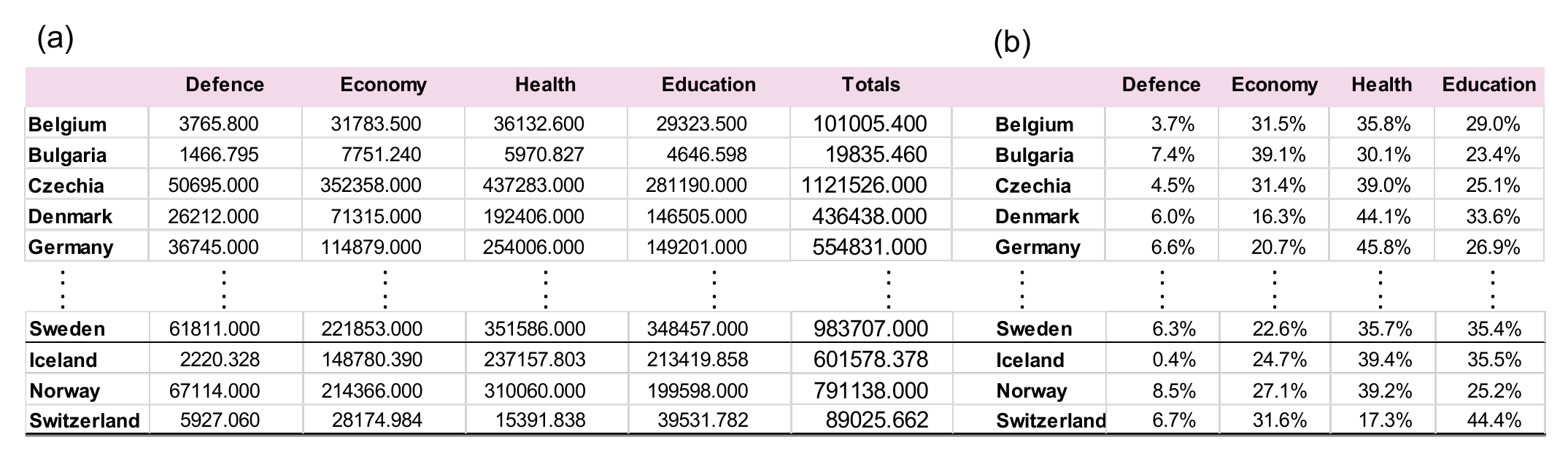}
%
% If no graphics program available, insert a blank space i.e. use
%\picplace{5cm}{2cm} % Give the correct figure height and width in cm
%
\caption{Expenditure in European Union countries plus Iceland, Norway and Switzerland on four budget items, in 2019. In (a) the amounts are in millions of local currency. In (b) the amounts are expressed as percentages relative to the totals.}
%\caption{If the width of the figure is less than 7.8 cm use the \texttt{sidecapion} command to flush the caption on the left side of the page. If the figure is positioned at the top of the page, align the sidecaption with the top of the figure -- to achieve this you simply need to use the optional argument \texttt{[t]} with the \texttt{sidecaption} command}
\label{Excel}       % Give a unique label
\end{figure}

Consider the table in Fig.~\hspace{-0.05cm}\ref{Excel}(a): these are amounts spent in 2019 on four different budget items in the European Union (EU), showing only six EU countries with the addition of Iceland, Norway and Switzerland.
The full data set consists of 30 countries.
The amounts are in millions of local currency and are clearly not comparable across the  countries.
Converting all the amounts to the same currency, for example euros, alleviates but does not solve the data coding problem, since some countries are small and others are large.
Seeing that only relative amounts are of principal interest, it seems that comparability across countries is assured by simply expressing the four amounts in each row as percentages, as in Fig.~\hspace{-0.05cm}\ref{Excel}(b).
The rows of data in Fig.~\hspace{-0.05cm}\ref{Excel}(b) are called \textit{compositions}: nonnegative multivariate data with the constant sum constraint, 100\% in this case.
The act of transforming the original monetary values to proportions, or percentages, is called in this context \textit{normalization} or \textit{closure}.

But is the coding problem really solved?  
If additional budget items were added to the table in Fig.~\ref{Excel}(a), such as public services, social protection and culture, the relative amounts in Fig.~\ref{Excel}(b) would change, necessarily decreasing, and they would reduce by different proportions since the additional budget amounts would not be in proportion to the totals of the four budget items shown in Fig.~\ref{Excel}(a).
This is what makes compositional data different from any other multivariate data in Statistics --- the values of each component in the table depend on the values of the other components.
It would make no sense, for example, to compute correlations on such a data matrix, since there are necessarily many negative correlations created by the constant sum constraint, and the correlation between health and education would be different in Fig.~\ref{Excel}(b) from the correlation between health and education in an expanded table of budget items also expressed as compositions.

\citet{Aitchison:82,Aitchison:86} showed that using ratios of the components was a solution to the data coding issue. 
Ratios remain constant between two components irrespective of adding components to or removing components from a composition --- they are said to be \textit{subcompositionally coherent}.
Furthermore, Aitchison proposed that ratios be logarithmically transformed: for example, if $X_j$ and $X_k$ are two components then the \textit{logratio} transformation is $\log(X_j/X_k) = \log(X_j) - \log(X_k)$, i.e. the difference in their logarithms, which is a linear transformation on the log scale. 

The objective of this chapter is to express the theory of logratio transformations in linear algebra terms, and in the process give a flavour of the visualization possibilities of logratios and their interpretation.
In Section 2, basic definitions and results are given in matrix-vector form, as well as inverse logratio transformations, including the important topic of log-contrasts. 
Section 3 shows how logratios can be visualized through cluster analysis and biplots, and Section 4 concludes with a discussion.

\section{Basic algebraic definitions and results}
\label{sec:2}
% Always give a unique label
% and use \ref{<label>} for cross-references
% and \cite{<label>} for bibliographic references
% use \sectionmark{}
% to alter or adjust the section heading in the running head
The practical aspects of compositional data analysis are given in two recent books, by \citet{Greenacre:18} and \citet{Filzmoser:18}, and a comprehensive review is given by \citet{Greenacre:21}.
In this section, the algebra of compositional data analysis is given in matrix--vector notation.

\subsection{Logratio transformations and associated pattern matrices}

\vspace{-0.1cm}
Suppose $\bx$ ($J\times 1$) is a $J$-component compositional vector of positive values with sum 1: i.e. $\bone\tr\bx = 1$, where $\bone$ is a vector of ones of appropriate order, in this case a vector of $J$ ones.
If $\log(\bx)$ denotes the vector of log-transformed values, then almost all logratio transformations are defined by a linear transformation of the form $\bP \log(\bx)$, where $\bP$ is called the \textit{logratio pattern matrix}.
If $\bX$ ($I\times J)$ denotes the data set, with sampling units as rows and components as columns, then the compositions are in the rows and the constant row sums are defined by postmultiplication by $\bone$: $\bX {\bone} = {\bone}$, where the $\bone$ on the right is $J\times 1$.
Similarly, if $\bL = \log(\bX)$ denotes the $I\times J$ matrix of logarithms of $\bX$, then the application of the pattern matrix to the rows of $\bL$ implies post-multiplication on the right by the transpose of the pattern matrix: $\bL \bP\tr$.
In the following the matrix $\bP$ will be subscripted by the type of logratio transformation.

The pairwise logratio pattern matrix $\bP_\textsc{lr}$ ($\frac{1}{2}J(J-1) \times J$) is defined as:
\begin{equation}
  \bP_\textsc{lr} =
  \begin{bmatrix*}[r]
    \ 1  & -1  &  0  &  0  &  0 & \cdots &  0  &  \ \ 0  &  0  \ \\
    \ 1  &  0  & -1  &  0  &  0 & \cdots &  0  &  \ \ 0  &  0  \ \\
    \ 1  &  0  &  0  & -1  &  0 & \cdots &  0  &  \ \ 0  &  0  \ \\
%    1  &  0  &  0  &  0  & -1 & \cdots &  0  &  \ \ 0  &  0  \ \\
    \ \vdots & \vdots & \vdots  & \vdots & \vdots & \ddots & \vdots & \ \ \vdots & \vdots \  \\
    \ 1  &  0  &  0  &  0  &  0 & \cdots &  0  &  \ \ 0  & -1  \ \\
    \ 0  &  1  & -1  &  0  &  0 & \cdots &  0  &  \ \ 0  &  0  \ \\  
    \ 0  &  1  &  0  & -1  &  0 & \cdots &  0  &  \ \ 0  &  0  \ \\
    \ 0  &  1  &  0  &  0  & -1 & \cdots &  0  &  \ \ 0  &  0  \ \\  
    \ \vdots & \vdots & \vdots  & \vdots & \vdots & \ddots & \vdots & \ \ \vdots & \vdots \  \\ 
    \ 0  &  0  &  0  &  0  &  0 & \cdots &  1  &  \ \ 0  &  -1  \ \\      
    \ 0  &  0  &  0  &  0  &  0 & \cdots &  0  &  \ \ 1  & -1  \        
  \end{bmatrix*}
  \label{LRpattern}
\end{equation}
so that the $I\times \frac{1}{2}J(J-1)$ matrix of pairwise logratios is $\bL \bP_\textsc{lr}\tr$.
Each row of $\bP_\textsc{lr}$ defines a pairwise logratio (LR), when applied to the logarithm of a compositio, $\log(\bx)$.
Notice the lexicographic ordering in the rows, corresponding to the ratio pairs (12), (13), (14), $\cdots$, (1$J$), (23), (24), (25), $\cdots$, ($J$--2,$\,J$), ($J$--1,$\,J$).  
We will abbreviate the term ``pairwise logratio'' by LR throughout the rest of this chapter.

\begin{figure}[h!]
\sidecaption
\hspace{-0.4cm}
\includegraphics[scale=.37]{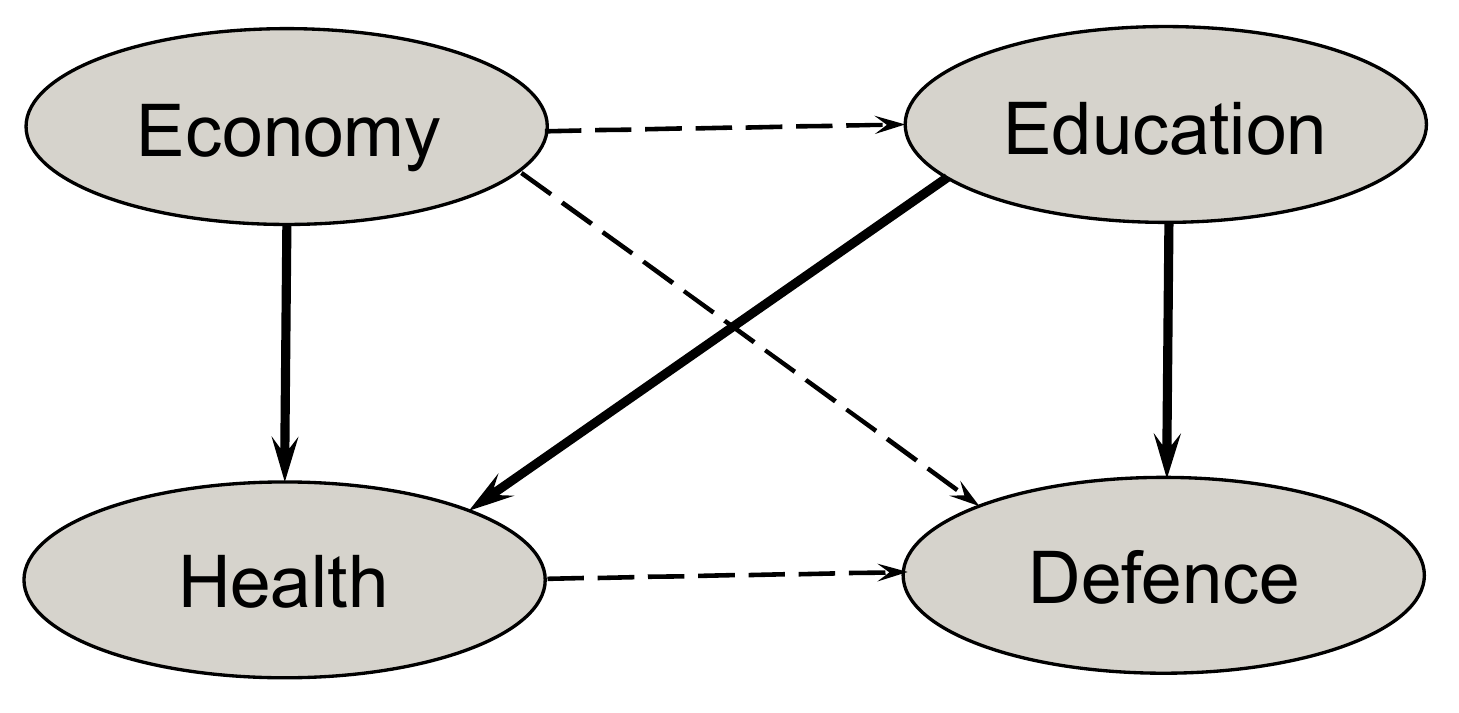}
%
% If no graphics program available, insert a blank space i.e. use
%\picplace{5cm}{2cm} % Give the correct figure height and width in cm
%
\caption{Directed acyclic graph (DAG) that connects the four components of Fig.~\ref{Excel}(a), indicated by the solid arrows. The dashed arrows indicate the other three pairwise logratios that can be obtained from those of the DAG.}
%\caption{If the width of the figure is less than 7.8 cm use the \texttt{sidecapion} command to flush the caption on the left side of the page. If the figure is positioned at the top of the page, align the sidecaption with the top of the figure -- to achieve this you simply need to use the optional argument \texttt{[t]} with the \texttt{sidecaption} command}
\label{DAG}       % Give a unique label
\end{figure}

The matrix of LRs is of rank $J-1$, assuming $I \geq J$, otherwise it is of rank $I-1$ (we shall assume for ease of description that there are at least as many rows as columns in the compositional data matrix).
This can be seen easily using a result by \citet{Greenacre:18,Greenacre:19} that a connected directed acyclic graph (DAG) consisting of $J-1$ LRs generates all the $\frac{1}{2}J(J-1)$ pairwise logratios through linear combinations.
For example, for the four-component example of Fig.~\ref{Excel}(a) with six LRs, a possible connected DAG is shown by the three solid arrows in Fig.~\ref{DAG}, representing the ratios Health/Economy, Health/Education and Defence/Education (the arrow points towards the numerator component). 
The other three logratios, indicated by dashed arrows, can be obtained from those in the DAG as ratios combined either through multiplication or division depending on the direction of the arrows: Education/Economy = (Health/Economy) / (Health/Education)), or in linearized logratio form using addition or subtraction: log(Education) -- log(Economy = [log(Health)--log(Economy)] -- [log(Health)--log(Education)].

\vspace{-0.5cm}
\hspace{-0.2cm}

\noindent
Assuming the four components above are in the order $\{$Defence, Economy, Health, Education$\}$ as in the table in Fig.~\ref{Excel}, the pattern matrix associated with the three logratios (solid arrows) in the DAG above has this form:
\begin{equation}
  \bP =
  \begin{bmatrix*}[r]
    \ 0  & -1  &  \ \ 1  &  0  \ \\
    \ 0  &  0  &  \ \ 1  & -1  \ \\
    \ 1  &  0  &  \ \ 0  & -1  \ 
  \end{bmatrix*}
  \label{LRpatternDAG}
\end{equation}

\begin{figure}[h!]
\sidecaption
\hspace{-0.4cm}
\includegraphics[scale=.37]{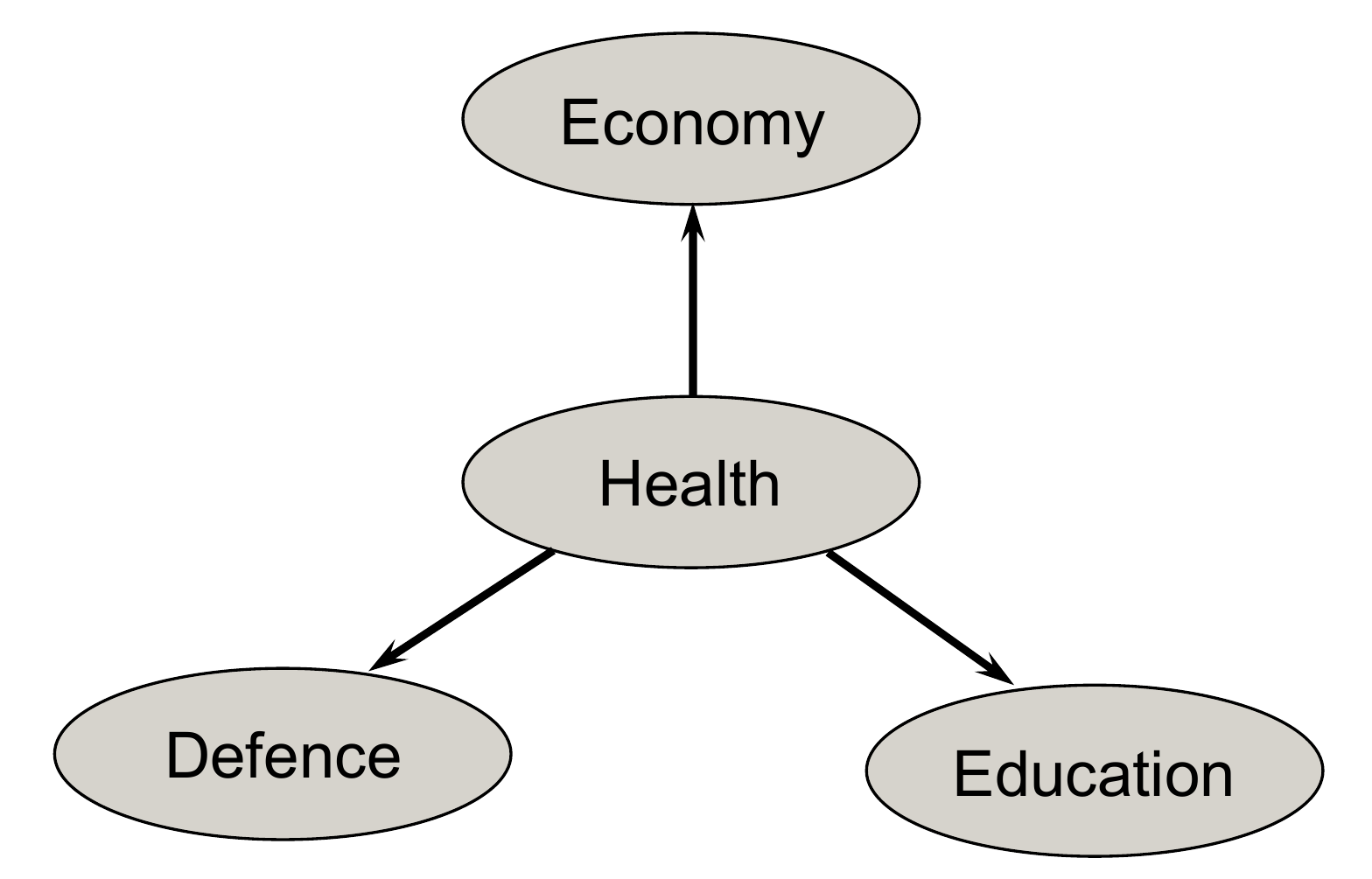}
%
% If no graphics program available, insert a blank space i.e. use
%\picplace{5cm}{2cm} % Give the correct figure height and width in cm
%
\caption{DAG corresponding to the ALR transformation of the four components in Fig.~\ref{Excel}, where Health is the reference part.}
%\caption{If the width of the figure is less than 7.8 cm use the \texttt{sidecapion} command to flush the caption on the left side of the page. If the figure is positioned at the top of the page, align the sidecaption with the top of the figure -- to achieve this you simply need to use the optional argument \texttt{[t]} with the \texttt{sidecaption} command}
\label{DAGalr}       % Give a unique label
\end{figure}

A special case of the LRs is the \textit{additive logratio} (ALR) transformation, the set of $J-1$ LRs where the denominator component is the same, called the {\it reference} component.
For example ,if the last component is the reference, then the ALR pattern matrix $\bP_\textsc{alr}$ ($(J-1)\times J$) is:
\begin{equation}
  \bP_\textsc{alr}  =
  \begin{bmatrix*}[r]
    \   1  & \   0   &  \  0   &  \cdots & \  0  & \ -1\ \\
    \   0  & \   1   &  \  0   &  \cdots & \  0  & \ -1\ \\
    \   0  & \   0   &  \  1   &  \cdots & \  0  & \ -1\ \\
    \ \vdots & \  \vdots  & \  \vdots   &  \ddots & \ \vdots & \ \vdots\ \\
    \   0  & \   0   &  \  0   &  \cdots  & \  1 & \ -1 \    
  \end{bmatrix*}
  \label{ALRpattern}
\end{equation}
The matrix of ALRs,  $\bL \bP_\textsc{alr}\tr$ $(I\times (J-1))$, is of rank $J-1$.
For example, if for the four components in Fig.~\ref{Excel} the third component Health was chosen as the reference part, then (\ref{ALRpattern}) would be a $3\times 4$ matrix with the $-1$s down the third column and a $1$ in each row in columns 1, 2 and 4.
Fig.~\ref{DAGalr} shows the DAG associated with this ALR transformation, where Health is placed in the centre and arrows emanate outwards to the other three components.

\hspace{-0.2cm}

The next most important logratio transformation is the \textit{centered logratio} (CLR) transformation, the ratio of each component divided by the geometric mean of all the components.
The usual unweighted definition is the following, for a row $[\,x_1, x_2,\ldots,x_J\,]$ of $\bX$:
\begin{equation}
  {\rm CLR}(j) = \log\Biggl( \frac{x_j}{{\bigl(}\prod_{k} x_{k}{\bigr)}^{1/J}} \Biggr) = \log(x_j)-\frac{1}{J} \sum_{k} \log(x_{k})   \quad j = 1,\ldots, J
  \label{CLR}
\end{equation}
but it is preferred here to give a more general weighted definition assuming positive weights $c_j$ ($j=1,\ldots,J$) for the components, where $\sum_j c_j = 1$, and thus a weighted geometric mean in the denominator:
\begin{equation}
  {\rm CLR}(j) = \log\Bigl( \frac{x_j}{\prod_{k} x_{k}^{c_k}} \Bigr) = \log(x_j)-\ \sum_{k} c_k \log(x_{k})   \quad j = 1,\ldots, J
  \label{CLRw}
\end{equation}
Hence, (\ref{CLR}) is the special case of (\ref{CLRw}) with equal weights $1/J$ for all the $J$ components.

The CLR pattern matrix $\bP_\textsc{clr}$ ($J \times J$) for the general case is:
\begin{equation}
  \bP_\textsc{clr}  =
  \begin{bmatrix*}
    1-c_1  &    -c_2   &    -c_3   &  \cdots &  -c_J  \ \\
     -c_1  &   1-c_2   &    -c_3   &  \cdots &  -c_J  \ \\
     -c_1  &    -c_2   &   1-c_3   &  \cdots &  -c_J  \ \\
    \vdots &  \vdots   &  \vdots   &  \ddots & \vdots \ \\
     -c_1  &    -c_2   &    -c_3   &  \cdots & 1-c_J  \    
  \end{bmatrix*}
  \label{CLRpattern}
\end{equation}
Notice that $\bP_\textsc{clr}$ is just the idempotent centering matrix $\bI - \bone\bc\tr$, where $\bc\tr = [\, c_1\ c_2 \cdots c_J\,]$.
The rows of $\bP_\textsc{clr}$ sum to 0, and the matrix of CLRs, $\bL \bP_\textsc{clr}\tr$, has rank $J-1$, just like the matrix of LRs, $\bL \bP_\textsc{lr}$ and the matrix of ALRs, $\bL \bP_\textsc{alr}$.

More complex logratio transformations are the \textit{isometric logratios} (ILRs) \citep{Egozcue:05} and their slightly simpler special case, the \textit{pivot logratios} (PLRs) \citep{Filzmoser:18} --- these are often called ``balances'' although the term can be misleading \citep{Greenacre:20, GreenacreGrunskyBaconShone:20}.
ILRs are log-transformed ratios of geometric means of groups of components.
In PLRs one of these groups is a single component and a linear independent set of $J-1$ PLRs takes the components in a fixed order and the numerator of each ratio is a single component and the denominator is the geometric mean of the others ``to the right of'' the numerator component.
For both ILRs and PLRs there is a scalar constant involved, which is omitted here for simplicity --- see Greenacre (2018) for the exact (unweighted) definition.
Again it is preferred to give the more general weighted definition here, which will be useful in practice when parts are considered differentially weighted.
The pattern matrix $\bP_\textsc{plr}$ ($(J-1) \times J)$ for the weighted case is:
\begin{equation}
  \bP_\textsc{plr}  =
  \begin{bmatrix*}
     \ 1    & \ \ \ -\frac{c_2}{c_2+\cdots+c_J}  & \ \ -\frac{c_3}{c_2+\cdots+c_J}  &  \cdots &   -\frac{c_{J-2}}{c_2+\cdots+c_J}  &  \ \ -\frac{c_{J-1}}{c_2+\cdots+c_J}  & \ \ -\frac{c_J}{c_2+\cdots+c_J}  \\[1.2ex]
     \ 0  &   \ \ \ 1   &   \ \  -\frac{c_3}{c_3+\cdots+c_J} & \cdots  &  -\frac{c_{J-2}}{c_3+\cdots+c_J}  &  \ \ -\frac{c_{J-1}}{c_3+\cdots+c_J}  &   \ \ -\frac{c_J}{c_3+\cdots+c_J} \\[1.2ex]
     \ 0  &  \ \  \ 0   &  \ \  1  &  \cdots & -\frac{c_{J-2}}{c_4+\cdots+c_J} & \ \ -\frac{c_{J-1}}{c_4+\cdots+c_J} & \ \  -\frac{c_J}{c_4+\cdots+c_J} \\[1.2ex]
    \vdots &  \ \ \ \vdots   &  \vdots   &  \ddots & \vdots & \ \ \vdots & \ \ \vdots \\[1.2ex]
     \ 0  &  \ \  \ 0   &  \ \  1  &  \cdots & 1  & -\frac{c_{J-1}}{c_{J-1}+c_J} & \ \  -\frac{c_J}{c_{J-1}+c_J} \\[1.2ex]
     \  0 &  \ \ \  0  &  \ \ 0  &  \cdots & 0 & 1 & \ \ -1    
  \end{bmatrix*}
  \label{PLRpattern}
\end{equation}

\begin{figure}[h!]
\sidecaption
\hspace{-0.4cm}
\includegraphics[scale=.32]{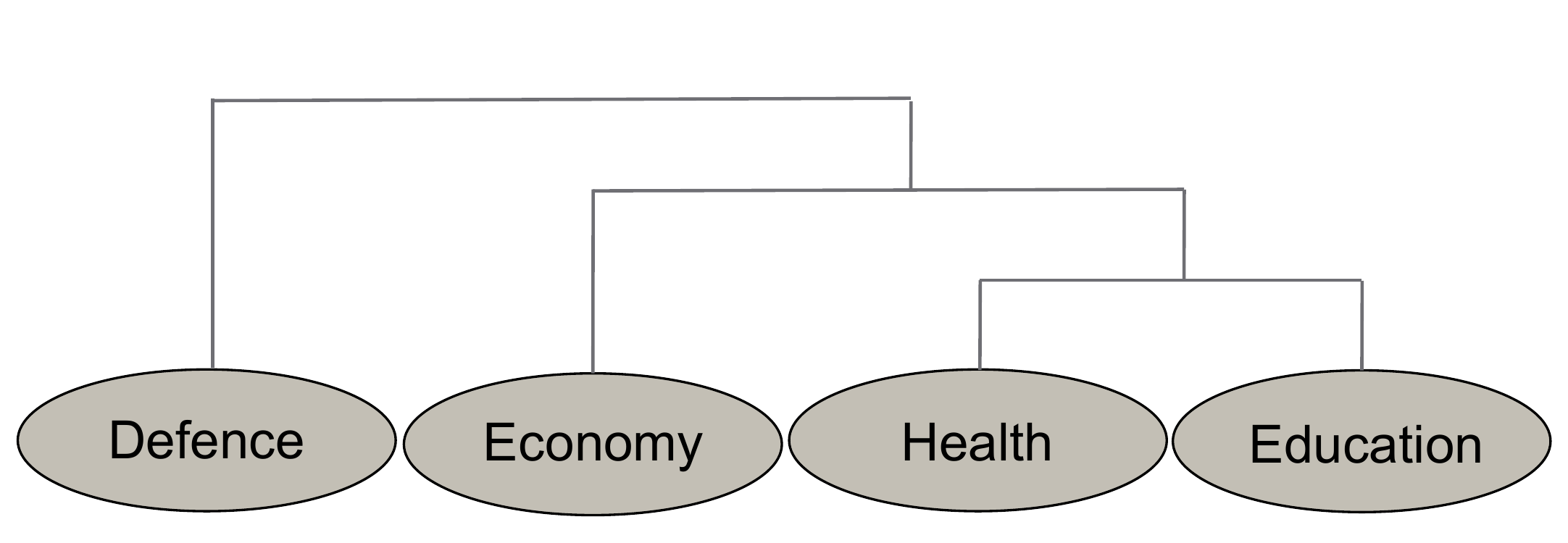}
%
% If no graphics program available, insert a blank space i.e. use
%\picplace{5cm}{2cm} % Give the correct figure height and width in cm
%
\caption{Dendrogram graph associated with three PLRs corresponding to the order of the components Defence, Economy, Health, Education. 
}
%\caption{If the width of the figure is less than 7.8 cm use the \texttt{sidecapion} command to flush the caption on the left side of the page. If the figure is positioned at the top of the page, align the sidecaption with the top of the figure -- to achieve this you simply need to use the optional argument \texttt{[t]} with the \texttt{sidecaption} command}
\label{PLRgraph}       % Give a unique label
\end{figure}

Like the pattern matrices before, (\ref{LRpattern}), (\ref{ALRpattern}) and (\ref{CLRpattern}), the row sums of $\bP_\textsc{plr}$ are zero and the matrix $\bL \bP_\textsc{plr}\tr$ of PLRs has rank $J-1$.
There are $J!$ permutations of the $J$ components, hence $J!$ sets of PLRs possible, depending on the ordering of the $J$ components.
For the four components of Fig.~\ref{Excel}, in the order given, a graph of the associated PLRs is in the form of the binary dendrogram of Fig.~\ref{PLRgraph}.
Since there are $4! = 24$ ordered permutations possible for this example, there are 24 different sets of PLRs possible.
Notice that the last member of a set of PLRs is a simple LR.
\vspace{-0.2cm}

\begin{figure}[h!]
\sidecaption
\hspace{-0.4cm}
\includegraphics[scale=.33]{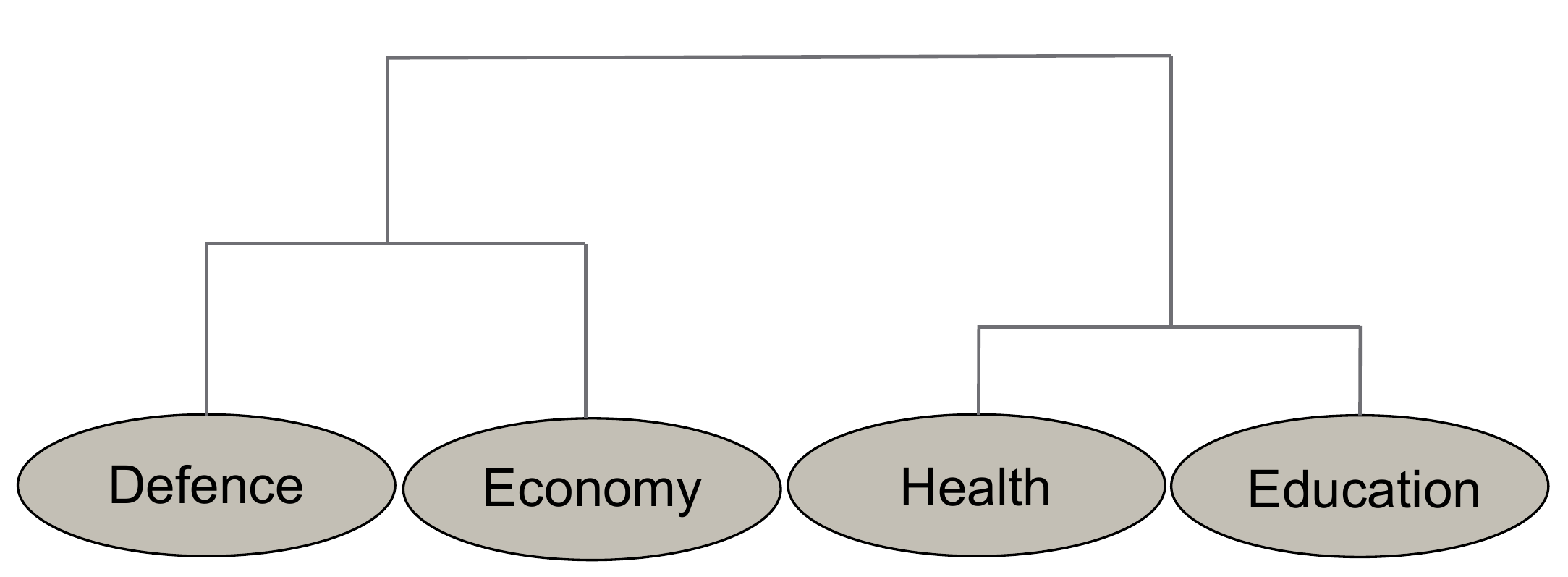}
%
% If no graphics program available, insert a blank space i.e. use
%\picplace{5cm}{2cm} % Give the correct figure height and width in cm
%
\caption{Dendrogram graph associated with three ILRs corresponding to the three ratios (moving down the tree): (1) geometric mean of Defence and Economy / geometric mean of Health and Education, (2) Defence / Economy, (3) Health\ /\ Education. 
}
%\caption{If the width of the figure is less than 7.8 cm use the \texttt{sidecapion} command to flush the caption on the left side of the page. If the figure is positioned at the top of the page, align the sidecaption with the top of the figure -- to achieve this you simply need to use the optional argument \texttt{[t]} with the \texttt{sidecaption} command}
\label{ILRgraph}       % Give a unique label
\end{figure}

The ILRs involving more general ratios of geometric means is usually defined according to a dendrogram graph, and the number of possible dendrograms increases even more rapidly with the number of components.
The ILR pattern matrix becomes more difficult to express in general, so we will just give the special case associated with the dendrogram in Fig.~\ref{ILRgraph}, assuming the order shown of the components.
\vspace{-0.2cm}

\noindent
The specific pattern $3\times 4$ matrix for this four-component example corresponding to Fig.~\ref{ILRgraph}, again with rows summing to 1, would thus be:
\begin{equation}
  \bP_{\sf ILR}  =
  \begin{bmatrix*}
     \frac{c_1}{c_1+c_2}     &  \ \ \frac{c_2}{c_1+c_2} & \ \ -\frac{c_3}{c_3+c_4} & \ \ -\frac{c_4}{c_3+c_4} \\[0.7ex]
     1    &  \ \ -1   &   \ \  \ \ 0    & \ \ \ \ 0 \\[0.7ex]
     0     &  \ \ \ \  0   &  \ \  \ \ 1 & \ \ -1       
  \end{bmatrix*}
  \label{ILRpattern}
\end{equation}

Finally, there is a class of nonlinear transformations called \textit{amalgamation} (or \textit{summated}) \textit{logratios}, abbreviated as SLRs.
Like the LRs, these are true balances between the components as they use sums rather than geometric means when combining components in the ratios \citep{Greenacre:20}.
For example, the dendrogram in Fig.~\ref{ILRgraph} would translate to the following three SLRs, the last two of which are regular LRs:
\begin{equation}
  (1) \log\left(\frac{\textrm{Defence+Economy}}{\textrm{Health+Education}}\right)\quad
  (2) \log\left(\frac{\textrm{Defence}}{\textrm{Economy}}\right) \quad
  (3) \log\left(\frac{\textrm{Health}}{\textrm{Education}}\right) 
  \label{SLRs}
\end{equation}
The first SLR above cannot be written as a linear function of the logarithms, but has the advantage of being more easily interpreted in a practical application.
It is still isomorphic, however, as shown in the next subsection.

\subsection{Inverting logratio transformations}
\label{subsec:2.2}
Each of the logratio transformations defined in the previous subsection can be inverted back to the original compositions, including the nonlinear SLRs.
Each in turn relies on a square \textit{inversion pattern matrix}, denoted by $\bQ$, closely related to the respective logratio pattern matrix $\bP$.
The back-transformation in each case involves solving a system of $J$ linear equations of the form $\bQ \bx = \be$, i.e. $\bx = \bQ^{-1} \be$, where $\bQ$ involves elements that are functions of the logratios, and $\be$ is the vector $[\,0\ \,0 \,\cdots\, 0\ \,1\,]\tr$. 

Starting with the ALRs, suppose that $\by = \bP_\textsc{alr} \bx$ is the vector of $J-1$ ALRs, where $\bx$ is a $J\times 1$ composition and $\bP_\textsc{alr}$ is given by (\ref{ALRpattern}).
Then the inverse operation of finding $\bx$  from $\by$ is the solution of the following equation:
\begin{equation}
     {\bQ}_\textsc{alr} \bx = 
     \begin{bmatrix} 
        \ 1 & \ 0 & \cdots & 0 & \ -e^{y_1} \\ \vspace{-0.1cm}
        \ 0 & \ 1 & \cdots & 0 & \ -e^{y_2} \\  
      \ \vdots & \ \vdots & \ddots & \vdots & \vdots \\ 
      \  0 & \ 0 & \cdots & 1 & \ -e^{y_{J-1}} \\
      \  1 & \ 1 & \cdots & 1 &  \ 1 
     \end{bmatrix}
     {\bf x} = 
     \begin{bmatrix} 
        0  \\ \vspace{-0.1cm}
        0  \\  
      \vdots  \\ 
        0  \\
        1  
     \end{bmatrix}
  \label{invALR2}
\end{equation}
Notice that the $-1$s in the pattern matrix, i.e. in the denominator positions, are substituted by $-e^{y_1}, -e^{y_2}, \ldots, -e^{y_{J-1}}$ in rows $1,2,\ldots,J-1$ respectively, and then a row of $1$s is added, which with the last $1$ in the right hand vector imposes the constraint that the $x_j$s in the solution sum to 1 \citep{Greenacre:20}.
This is equivalent to the simpler calculation of exponentiating the $J-1$ logratios and expanding them with a 1, i.e. $[\ e^{y_1}\ e^{y_2} \cdots \ e^{y_{J-1}}\ \ 1\,]$ and then normalizing the result to sum to 1.
So, although using the linear equations approach by solving (\ref{invALR2}) is actually an inefficient way of back-transforming the compositions, it is enlightening because it provides a way to deal with any set of $J-1$ independent LRs, e.g. the three LRs defined in the DAG of Fig.~\ref{DAG}.

The way of inverting a set of $J-1$ independent pairwise logratios (LRs) is a simple generalization of the system of equations (\ref{invALR2}) above.
The $(J-1)\times J$ pattern matrix ${\bf P}_\textsc{lr}$ for the transformation to the LRs, now has each of the $J-1$ rows  corresponding to a specific LR, $\log(x_j/x_{j^\prime})$, with a $1$ in column $j$ of the numerator part and $-1$ in column $j^\prime$ of the denominator part.
To obtain the inversion pattern matrix, the $-1$ is again replaced in row $k$ by  $-e^{y_{k}}$ and then a row of $1s$ is added as the last row, as before.
For example, for the DAG in Fig.~\ref{DAG}, the logratio pattern matrix and corresponding inversion pattern matrix, with their matrix equations are, in the order Defence, Economy, Health, Education, and LRs Health/Economy, Health/Education and Defence/Education, for $\bx$ ($4\times 1$) a compositional vector, $\by$ ($3\times 1$) the logratio transformation, and $\be = [\,0\, \ 0\,\ 0\,\ 1\,]\tr$:
\begin{equation}
  \bP_\textsc{lr} =
  \begin{bmatrix*}[r]
    0 & -1 & \ 1 &  0 \\
    0 &  0 & \ 1 & -1 \\
    1 &  0 & \ 0 & -1 
  \end{bmatrix*},
  \ \  {\rm i.e.\ } \by = \bP_\textsc{lr}\log(\bx) 
  \quad
  \bQ_\textsc{lr} =
  \begin{bmatrix}
    0 & -e^{y_1} & 1 &  0 \\
    0 &  0 & 1 & -e^{y_2} \\
    1 &  0 & 0 & -e^{y_3} \\
    1 &  1  &  1  &  1
  \end{bmatrix}  ,
   \ \ {\rm i.e.\ } \bx = \bQ_\textsc{lr}^{-1} \be 
  \label{invDAG}
\end{equation}

The same approach can be used to invert a set of $J-1$ independent amalgamation logratio balances (SLRs), which must involve each part at least once.
The SLRs have a pattern matrix ${\bf P}_\textsc{slr}$ indicating which parts are in the numerator and in the denominator, but this is not the matrix of the transformation, which is not linear in $\log({\bf x})$, hence not a contrast matrix.
For example, the pattern matrix for the SLRs in (\ref{SLRs}) as well as the corresponding inversion pattern matrix are as follows:
\begin{equation}
  \bP_\textsc{slr} =
  \begin{bmatrix*}[r]
    1 &  1 & -1 &  -1 \\
    1 &  -1 & 0 &  0 \\
    0 &  0 &  1 & -1 
  \end{bmatrix*}
  \qquad
  \bQ_\textsc{slr} =
  \begin{bmatrix}
    1 & 1 & -e^{y_1} & -e^{y_1} \\
    1 & -e^{y_2} & 0 & 0  \\
    0 &  0 & 1 & -e^{y_3} \\
    1 &  1  &  1  &  1
  \end{bmatrix}  
  \label{invSLRs}
\end{equation}
Each row of ${\bf P}_\textsc{slr}$ corresponding to an SLR has $1$s in the columns of the numerator parts, and $-1$s in the columns of the denominator parts.
Then, as before, replace all the $-1$s in the $k$-th row of $\bQ_\textsc{slr}$ by $-e^{y_{k}}$ and add a row of $1s$ as the last row, which again automatically closes the parts in the solution.
The compositional vector can then be recovered by $\bx = \bQ_\textsc{lsr}^{-1} \be$. 
Notice that both the LR and ALR inverse transformations are special cases of the SLR one, where LRs and ALRs have only one numerator and one denominator part.
For specific examples of inverse transforms of sets of LRs and sets of SLRs, see  \cite{Greenacre:20}.

%For inverting the PLRs and more general ILRs, a slightly different approach can be adopted.
%For example, for the PLR pattern matrix $\bP_\textsc{plr}$ in (\ref{PLRpattern}), augment it with a row of ones:
%\begin{equation*}
%  {\bP^\ast} = 
%  \begin{bmatrix}
%    {\bP}_\textsc{plr} \\
%    {\bone}\tr 
%  \end{bmatrix}
%\end{equation*}
%Applying ${\bf P}^\ast$ to the logarithms of the composition, ${\bf P}^\ast\log(\bx)$ gives the $J-1$ PLRs in $\by$ and a $J$-th element $\bone\tr\log(\bx)$ equal to $I$ times the respective mean of the log-transformed composition: 
%\begin{equation*}
%  \by^\ast = 
%  \begin{bmatrix}
%    \by \\
%    \bone\tr\log(\bx)
%  \end{bmatrix}
%\end{equation*}
%Then $\log(\bx)$ can be recovered directly as $\log(\bx) = (\bP^\ast)^{-1} \by^\ast$ and exponentiating gives $\bx$ in normalized form.

\subsection{Log-contrasts}

A log-contrast is a linear combination of logarithms of all the components of composition, with the condition that the coefficients sum to $0$:
\begin{equation}
 \sum_j a_j \log(x_j), \quad {\rm where\ } \sum_j a_j = 0. 
  \label{logcontrast}
\end{equation}
The logratio pattern matrices ${\bf P}_\textsc{alr}$ in (\ref{ALRpattern}), ${\bf P}_\textsc{clr}$ in (\ref{CLRpattern}), and ${\bf P}_\textsc{plr}$ in (\ref{PLRpattern}), as well as a particular pattern matrix associated with a DAG such as (\ref{LRpatternDAG}), transform the vector $\log({\bf x})$ to the corresponding set of $J-1$ logratios, or $J$ logratios in the case of the CLR transformation.
For any one of these sets of logratios, denoted in general by $\ell_1, \ell_2,\ldots$, the coefficients $\bf c$ of a linear combination of them $c_1\ell_1 + c_2\ell_2 + \cdots $ can be converted to the coefficients of the log-contrast simply by pre-multiplying $\bf c$ by the transpose of the pattern matrix.
For example, for a linear combination of ALRs, the $J$ coefficients of the log-contrast are ${\bf a} = {\bf P}_\textsc{alr}\tr {\bf c}$.
This result is useful when a linear combination of logratios, used as explanatory variables in a generalized linear model, is estimated in explaining/predicting a response variable \citep{Coenders:20}.

This result is illustrated for four different transformations, for the linear modelling of a response variable in the form of the proportion of total budget in the same 30 countries devoted to Housing and Community Amenities, logarithmically transformed, denoted by $\log(y)$.
For example, if one defines the ALR with Health as the reference component, then the estimated regression model is, along with p-values in parentheses and proportion of explained variance, $R^2$:

\vspace{-0.2cm}
\small
\begin{equation*}
\hspace{-0.9cm}
\log(y) = -0.482 + 0.029\log\left(\frac{\textrm{\scriptsize Defence}}{\textrm{\scriptsize Health }}\right) + 1.032\log\left(\frac{\textrm{\scriptsize Economy}}{\textrm{\scriptsize Health}}\right) - 0.904\log\left(\frac{\textrm{\scriptsize Education}}{\textrm{\scriptsize Health}}\right) 
\end{equation*}
\vspace{-0.3cm}
\begin{equation*}
\hspace{1.9cm} (p = 0.004) \hspace{1.2cm} (p = 0.030) \hspace{1.2cm} (p = 0.83) \hspace{1cm} R^2 =  0.295
\end{equation*}
\normalsize

The coefficients of the log-contrast, using the corresponding ALR pattern matrix, are then  (always remembering the order of the components):
\begin{equation*}
{\bf a} = {\bf P}_\textsc{alr}\tr {\bf c} = 
\begin{bmatrix*}[r]
    1 &  0 &  0 \ \\
    0 &  1 &  0 \ \\
   -1 & -1 & -1 \ \\
    0 &  0 &  1 \ 
\end{bmatrix*} 
\begin{bmatrix*}[r]
    0.029 \ \\
    1.032 \ \\
    -0.904 \ 
\end{bmatrix*} =
\begin{bmatrix*}[r]
0.029 \ \\
1.032 \ \\
-0.157 \ \\
-0.904 \ 
\end{bmatrix*} 
\end{equation*}
It can be verified that $\bone\tr \ba = 1$.
Thus the regression model can be written as the constant plus the log-contrast:
\small
\begin{equation*}
\log(y) = -0.482 + 0.029 \log(\textrm{Defence}) + 1.032 \log(\textrm{Economy}) 
- 0.157 \log(\textrm{Health}) - 0.904 \log(\textrm{Education}) 
\end{equation*}
\normalsize

From the form of ${\bf P}_\textsc{alr}\tr$ and the fact that any ALR transformation will give the same log-contrast, it is clear that the coefficient of the reference part is the one that will change. For example, if Education is the reference, the regression coefficients turn out to be $[\ 0.029\ \ 1.032\ \ -0.157\ ]\tr$ and the log-contrast is shown to be identical: 
\begin{equation*}
{\bf a} = {\bf P}_\textsc{alr}\tr {\bf c} = 
\begin{bmatrix*}[r]
    1 &  0 &  0 \ \\
    0 &  1 &  0 \ \\
    0 &  0 &  1 \ \\
   -1 & -1 & -1 \ 
\end{bmatrix*} 
\begin{bmatrix*}[r]
    0.029 \ \\
    1.032 \ \\
    -0.157 \ 
\end{bmatrix*} =
\begin{bmatrix*}[r]
0.029 \ \\
1.032 \  \\
-0.157 \ \\
-0.904 \ 
\end{bmatrix*} 
\end{equation*}
For any set of pairwise logratios, once again the same log-contrast is obtained. For example, here is the regression model using the logratios in the DAG of Fig.~\ref{DAG}:

\vspace{-0.2cm}
\small
\begin{equation*}
\hspace{-0.7cm}
\log(y) = -0.482 - 1.032 \log\left(\frac{\textrm{\scriptsize Health}}{\textrm{\scriptsize Economy }}\right) + 0.875 \log\left(\frac{\textrm{\scriptsize Health}}{\textrm{\scriptsize Education}}\right) + 0.029 \log\left(\frac{\textrm{\scriptsize Defence}}{\textrm{\scriptsize Education}}\right) 
\end{equation*}
\vspace{-0.3cm}
\begin{equation*}
\hspace{2.1cm} (p = 0.004) \hspace{1.2cm} (p = 0.036) \hspace{1.4cm} (p = 0.83) \hspace{0.8cm} R^2 =  0.295
\end{equation*}
\normalsize

\noindent
Then, using the pattern matrix $\bP$ in (\ref{LRpatternDAG}):
\begin{equation*}
{\bf a} = {\bf P}\tr {\bf c} = 
\begin{bmatrix*}[r]
    0 &  0 &  1 \ \\
   -1 &  0 &  0 \ \\
    1 &  1 &  0 \ \\
    0 & -1 & -1 \ 
\end{bmatrix*} 
\begin{bmatrix*}[r]
    -1.032 \ \\
     0.875 \ \\
     0.029 \ 
\end{bmatrix*} =
\begin{bmatrix*}[r]
0.029 \ \\
1.032 \  \\
-0.157 \ \\
-0.904 \ 
\end{bmatrix*} 
\end{equation*}

The same result is obtained for the CLRs, as well as any set of ILRs or PLRs.
One difference with the CLRs is that the pattern matrix is $4\times 4$ and only three  coefficients are obtained in a regression, so the fourth one has to be set to zero.
In all cases the constant as well as the $R^2$ and the p-value for the whole model (which is $p=0.026$) are identical across the variations, as they all reduce to the same log-contrast.

Often, some type of variable selection is made to arrive at a more parsimonious model. 
Selecting fewer explanatory variables implies forming a subcomposition of the parts. 
A statistical criterion is needed to make the selection and there are many possible ways to do achieve this.
For example, one could do a permutation test on the coefficients of the log-contrast.
Using the CLR transformation, and randomizing the order of the response variable 999 times, the p-values for each log-contrast coefficient were estimated as:

\smallskip
\small
\centerline{Defence: $p=0.86$,\quad Economy: $p=0.003$, \quad Health: $p=0.66$, \quad Education: $p=0.037$}.
\normalsize

It seems that only one logratio, that of Economy/Education, can be used as a predictor, and gives the following result:

\vspace{-0.2cm}
\small
\begin{equation*}
\hspace{-1.2cm}
\log(y) = -0.562 + 1.010 \log\left(\frac{\textrm{\scriptsize Economy}}{\textrm{\scriptsize Eduction }}\right)
\end{equation*}
\vspace{-0.4cm}
\begin{equation*}
\hspace{2cm} (p = 0.002) \hspace{1.2cm} R^2 =  0.286
\end{equation*}
\normalsize
This leads to the trivial log-contrast of the two components in the model:
\small
\begin{equation*}
\log(y) = -0.562 + 1.010 \log(\textrm{Economy}) -1.010 \log(\textrm{Education})
\end{equation*}
\normalsize

\noindent
For more details about logratios used as predictors in linear modelling, see \cite{Coenders:20}.

\section{Logratio visualization}
In this section we look at various ways of visualizing a compositional data set.
Basically, once a logratio transformation is made, any of the various well-known multivariate visualization methods can be implemented, such as cluster analysis and dimension-reduced component methods.
Care has to be taken in the interpretation because of the unit sum constraint on the original data.
Since these methods rely on interpoint distances, the first thing to do is to define the distance measures between rows and between columns of the data matrix.

If the matrix ${\bf Z} = [z_{i,jj^\prime}]$ ($I\times \frac{1}{2}J(J-1)$) denotes the matrix of LRs $\log(x_{ij}/x_{ij^\prime})$, and ${\bf Y} = [y_{ij}]$ ($I\times J$) the CLR-transformed data set, then the \emph{logratio distance} $d_{ii^\prime}$ between samples $i$ and $i^\prime$, can be defined in two equivalent forms, shown for the weighted and unweighted versions in (\ref{dist_samples_weighted}) and (\ref{dist_samples_unweighted}) respectively \citep{Greenacre:18}:
 \begin{equation}
 d_{ii^\prime} = \sqrt{ {\sum\sum}_{j<j^\prime} c_j c_{j^\prime}(z_{i,jj^\prime} - z_{i^\prime ,jj^\prime})^2 } = \sqrt{\sum_j c_j (y_{ij} - y_{ij^\prime} )^2 }
  \label{dist_samples_weighted}
\end{equation}
 \begin{equation}
 d_{ii^\prime} = \sqrt{\frac{1}{J^2} {\sum\sum}_{j<j^\prime} (z_{i,jj^\prime} - z_{i^\prime ,jj^\prime} )^2 } = \sqrt{\frac{1}{J} {\sum}_j (y_{ij} - y_{ij^\prime} )^2  } 
  \label{dist_samples_unweighted}
\end{equation}
The advantage of the versions using CLRs is the use of a much narrower matrix, but --- as will be emphasized repeatedly --- the results should always be interpreted in terms of pairwise logratios.
The CLRs as such have no inherent interpretation as representing the components and simply act as a short cut to analysing all the pairwise logratios.

There are similar results for the columns, by transposing the data set, renormalizing and performing the same operations.
The samples are usually unweighted (i.e., with weights $1/I$) but can also be differentially weighted if required --- see \cite{Greenacre:18}. 
Later for the definition of the biplot the theory is presented in complete generality with weights on the rows and the columns.

\begin{figure}[h!]
\hspace{-0.2cm}
\includegraphics[scale=.42]{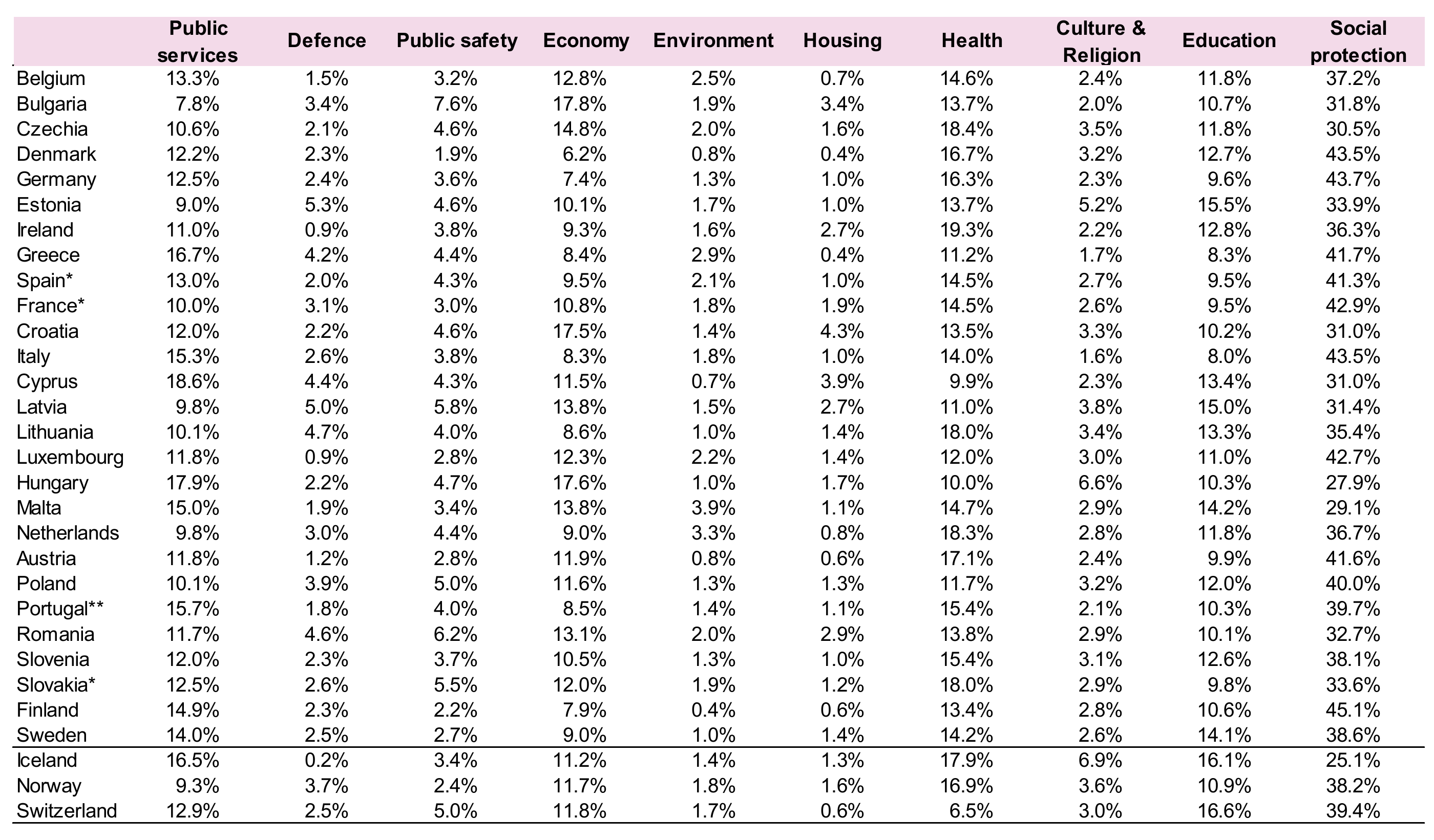}
\caption{Expenditure in European Union countries plus Iceland, Norway and Switzerland on ten budget items, in 2019. The data are expressed as percentages of the expenditure on these items (i.e., row sums are 100\%). Some columns names have been slightly abbreviated --- see the original longer names in Fig.~\ref{Clusterings} below.}
\label{Excel2}       % Give a unique label
\end{figure}

\begin{figure}[h!]
\hspace{-0.2cm}
\includegraphics[scale=.46]{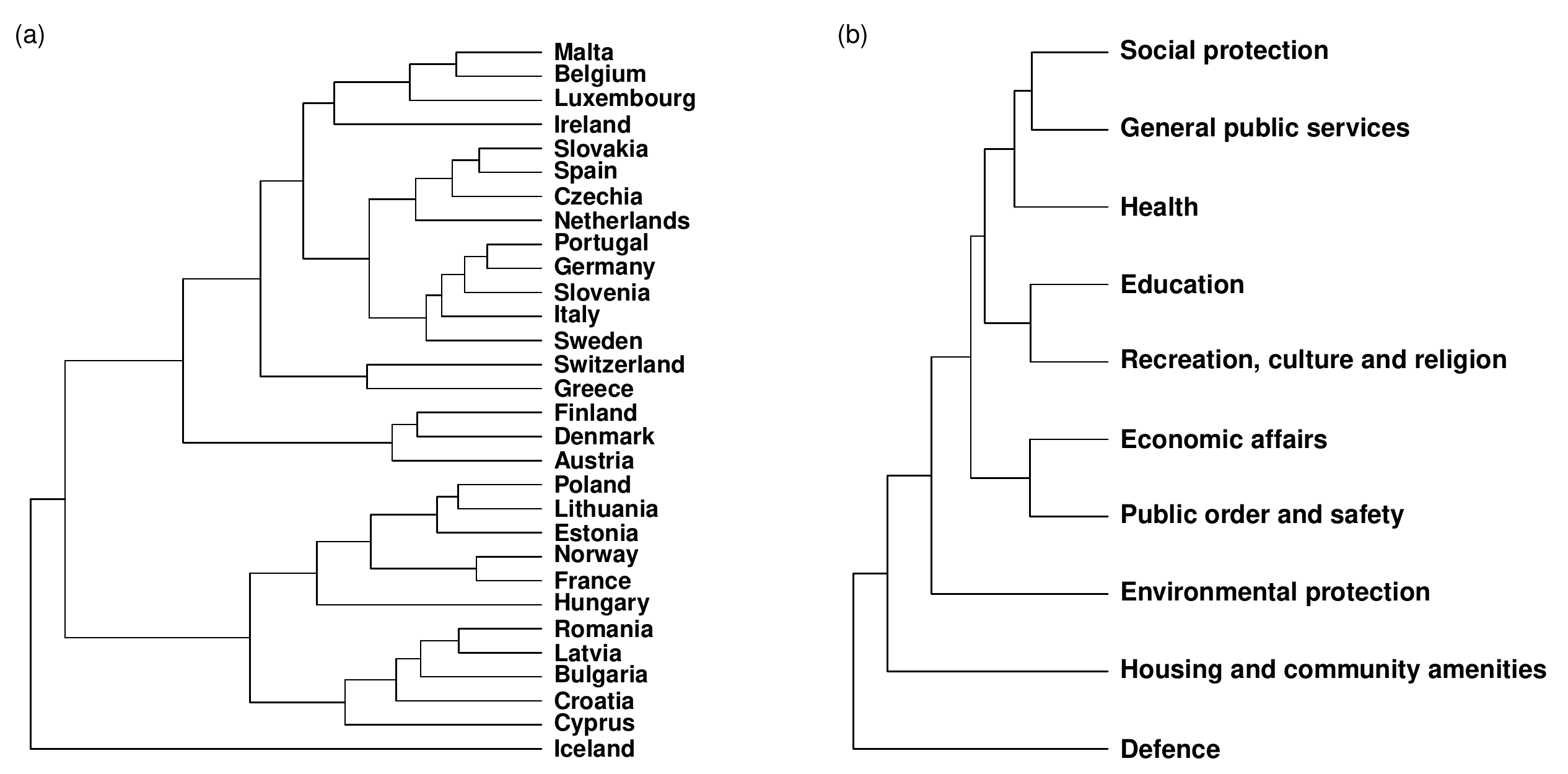}
\caption{Clustering of (a) countries and (b) budget items, using Ward clustering of the logratio distances. The budget items are weighted, as in (\ref{dist_samples_weighted}), using weights proportional to their average marginal proportions.}
\label{Clusterings}       % Give a unique label
\end{figure}

Fig.~\ref{Clusterings} shows the clustering the rows and columns respectively of an extended data set of country budget items (Fig.~\ref{Excel2}), using the \texttt{easyCODA} package \citep{Greenacre:18} in \textsf{R} \citep{R:21}, and Ward clustering in each case \citep{Ward:63}.

\begin{figure}[t!]
\hspace{-0.2cm}
\includegraphics[scale=.46]{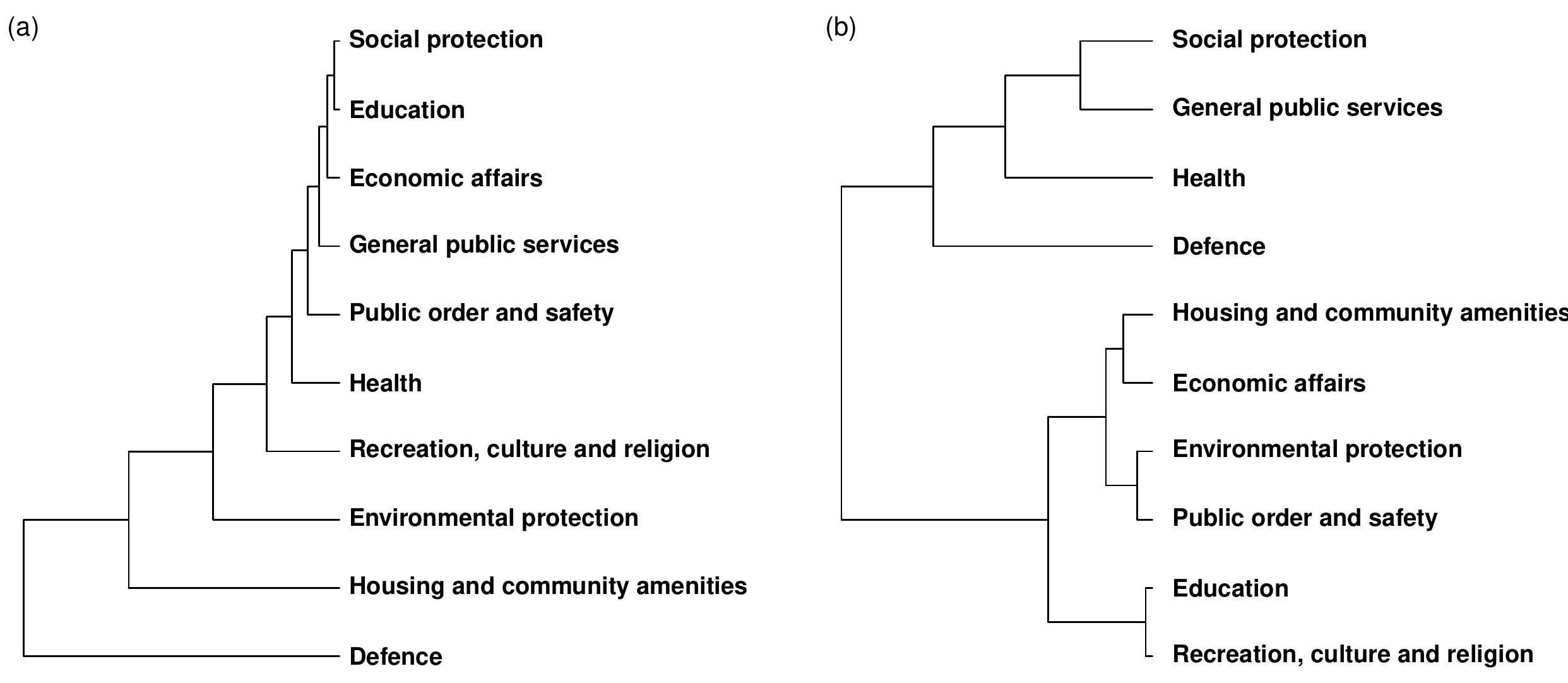}
\caption{Clustering of budget items using amalgamation clustering: (a) unweighted and (b) weighted.}
\label{Clusterings2}       % Give a unique label
\end{figure}

A completely different way of performing the clustering is by successively amalgamating the components \citep{Greenacre:20}, shown in Fig.~\ref{Clusterings2}.
Both unweighted and weighted versions are shown, giving different results, and different from the Ward clustering in Fig.~\ref{Clusterings}(b).
The unweighted version is identical to the graph structure of a set of PLRs, while the weighted version de-emphasizes the role of Defence, and shows that there are three pairs of items that could be merged: Housing and community amenities with Economic affairs, Environmental protection with Public order and safety, and Education with Recreation, culture and religion.

\begin{figure}[h!]
\hspace{-0.2cm}
\includegraphics[scale=.6]{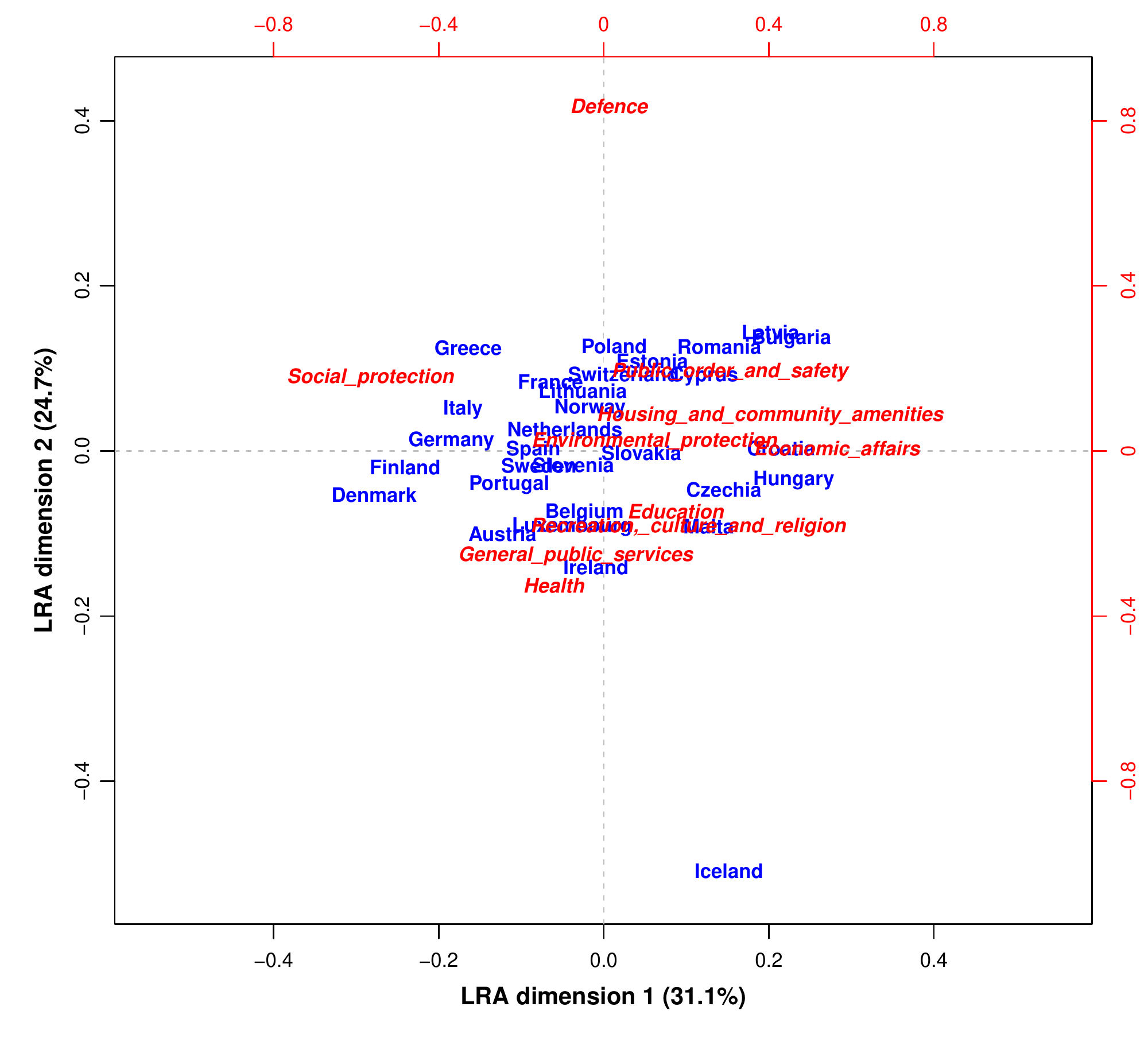}
\caption{Weighted logratio biplot of the data in Fig.~\ref{Excel2}, with asymmetric biplot scaling: rows in principal coordinates, columns in standard coordinates.}
\label{Biplot}       % Give a unique label
\end{figure}

Next, a logratio biplot is shown in Fig.~\ref{Biplot}, using logratio analysis (LRA).
This biplot is obtained using the singular value decomposition (SVD) of the double-centered matrix of the log-transformed compositional data set, $\log(\bX)$.
An unweighted or weighted version of LRA is possible, and here weighted LRA is used, where weights are imposed on the budget items equal to the average proportions of the items across the countries \citep{GreenacreLewi:09, Greenacre:18}.
This weighting will reduce the influence of some budget items with low average proportions but high logratio variance.
The configuration of the countries is approximating the weighted logratio distances in (\ref{dist_samples_weighted}).

Using matrix-vector notation, the sequence of steps to perform LRA and arrive at the biplot in Fig.~\ref{Biplot} is as follows, starting from the matrix of log-transformed compositional data, $\log({\bf X})$.
Note that the most general row- and column-weighted version is given here, with row and column weights $\br$ and $\bc$ respectively.
Usually, but not necessarily, the rows are equally weighted: $\br = (1/I)\bone$, and the description weighted or unweighted LRA refers to the nature of the column weights $\bc$, which can be different (e.g., by default equal to the marginal average proportions of the components \citep{GreenacreLewi:09}) or equal: $\bc = (1/J)\bone$ \citep{AitchisonGreenacre:02}.
%\interfootnotelinepenalty=10000
\begin{flalign}
& \textrm{Double-centre the matrix} \log({\bf X}): & \ & \hspace{-0.6cm} \mkern-1mu \bZ \mkern-2mu = ({\bf I} - {\bf 1r}\tr) \log({\bf X}) ({\bf I} - {\bf 1c}\tr)\tr & \label{Doublecenter} \\[3pt]
%\footnote{\samepage Notice that this is equivalent to column-centring the matrix {\bf Y} of CLRs, which is already row-centred:  $\widetilde{{\bf Y}} \mkern-2mu = ({\bf I} - {\bf 1r}\tr){\bf Y}$.}
%\vspace{-1cm}
%\end{flalign}
%\footnotetext{Notice that this is equivalent to column-centring the matrix {\bf Y} of CLRs, which is already row-centred:  $\widetilde{{\bf Y}} \mkern-2mu = ({\bf I} - {\bf 1r}\tr){\bf Y}$, where 
%${\bf Y} = \log({\bf X})({\bf I}-{\bf 1 c}\tr)\tr = \log({\bf X})({\bf I}-{\bf c 1}\tr)$
% --- see (A.7).}
%\vspace{-0.6cm}
%\begin{flalign}
& \textrm{Apply weights to rows and columns:}  & \ & \hspace{-0.6cm}
{\bf S}  = {\bf D}_r^\frac{1}{2} \bZ {\bf D}_c^\frac{1}{2} & \label{Weighting}\\[3pt]
& \textrm{Perform the SVD:} & \ & \hspace{-0.6cm}
{\bf S}  = {\bf UD}_\alpha{\bf V}\tr & \\[3pt]
& \textrm{Principal coordinates of rows:}  & \ & \hspace{-0.6cm}
{\bf F}  = {\bf D}_r^{\!-\frac{1}{2}}{\bf UD}_\alpha & \label{Rowpcoord}\\[3pt]
%& \textrm{Principal axes:}  & \ & \hspace{-0.6cm}
%{\bf A}  = {\bf D}_c^{\frac{1}{2}}{\bf V} & \\[3pt]
& \textrm{Standard coordinates of columns:} & \ & \hspace{-0.6cm}
\bGam  = {\bf D}_c^{\!-\frac{1}{2}}{\bf V} &\\[3pt]
& \textrm{Contribution coordinates of columns:}  & \ & \hspace{-0.6cm}
\bGam^{\mkern-1mu\ast} \mkern-7mu = {\bf D}_c^{\frac{1}{2}}\bGam = {\bf V} & \label{Contribution}
\end{flalign}

Two-dimensional biplots of the results are given by plotting the row principal coordinates in the first two columns of $\bf F$ with either the corresponding column standard coordinates in the first two columns of $\bGam$ (\textit{asymmetric biplot}) or those of the contribution coordinates in $\bGam^{\mkern-1mu\ast}$ (\textit{contribution biplot}).

The two-dimensional \textit{symmetric map} of the LRA is the plotting of the first two columns of the row principal coordinates {\bf F} in (\ref{Rowpcoord}) jointly with those of the column principal coordinates {\bf G}:
\vspace{-0.2cm}
\begin{flalign}
& \textrm{Principal coordinates of columns:}  & \ & \hspace{-2.2cm}
{\bf G}  = {\bf D}_c^{\!-\frac{1}{2}}{\bf VD}_\alpha &
\end{flalign}
This is not a biplot, strictly speaking, but has the practical advantage that the row and column points are equally scaled along the principal axes, their weighted variances both being equal to the amount of variance explained on the axes, i.e.~the eigenvalues or squared singular values {\small$\alpha$}$_k^2$ on axis $k$, $k=1,2$.

The sum of the squared singular values, $\sum_k ${\small$\alpha$}$_k^2$ (i.e.~sum of eigenvalues), equals the total logratio variance and {\small$\alpha$}$_k^2$ is the part of variance explained by axis $k$, usually expressed as a percentage of this total, as shown on the axes in Fig.~\ref{Biplot}. 

The steps (\ref{Doublecenter})--(\ref{Contribution}) are equivalent to performing a principal component analysis (PCA) on the matrix of (weighted) CLRs, because the CLRs are the row-centered $\log(\bX)$ (i.e., $\log({\bf X}) ({\bf I} - {\bf 1c}\tr)\tr$ in (\ref{Doublecenter})) and PCA automatically performs column-centering (i.e., the centering $({\bf I} - {\bf 1r}\tr)$ in (\ref{Doublecenter}), with equal weights in $\br$), hence the double-centering. The steps from (\ref{Weighting}) onwards define the PCA with its variations of display coordinates.
Because of the double-centering of $\log({\bf X})$, the definition of each principal component as a linear combination of the CLRs turns out to be a log-contrast.

The interpretation of Fig.~\ref{Biplot} is not the same as a regular PCA, however. 
The CLRs as variables are not interpretable \textit{per se}, but rather the differences between pairs of CLR points, which depict the pairwise logratios themselves.
\cite{AitchisonGreenacre:02} show that the LRA biplot optimizes the display of these pairwise logratios, which is not the case in a regular PCA (i.e., in a regular PCA the optimization of the variables is not equivalent to the optimization of the differences between pairs of variables).
Thus, to the horizontal dispersion in Fig.~\ref{Biplot} is due to logratios such as Public Order and Safety divided by Social Protection, while the vertical dispersion is dominated by the logratio of Defence relative to Health, for example, where it is clear that Iceland's ratios of Defence relative to the other budget items are low.

The equivalence between the PCA of the CLRs and the PCA of the LRs (i.e., LRA in both cases), can be neatly shown using the respective logratio pattern matrices defined earlier.
The proof follows the one given for unweighted logratios in the Appendix of \cite{AitchisonGreenacre:02}, but is more elegantly defined for the general weighted case.
The double-centred matrix $\bZ$ in (\ref{Doublecenter}) is the matrix $\bL =\log(\bX)$ post-multiplied by the transposed column-centering matrix, which is identical to the transposed CLR pattern matrix in (\ref{CLRpattern}), and pre-multiplied by the row-centering matrix, i.e. $\bZ = (\bI - \bone\br\tr) \bL \bP_\textsc{clr}\tr$.
The corresponding result for the matrix of pairwise LRs $\bL\bP_\textsc{lr}\tr$, is the row-centered matrix $\bY = (\bI - \bone\br\tr) \bL \bP_\textsc{lr}\tr$. 
The weights in the two respectve cases are $c_1, c_2, \ldots, c_J$ for the $J$ CLRs and $c_1c_2, c_1c_3, \ldots, c_{J-1}c_J$ for the $\frac{1}{2}J(J-1)$ LRs, gathered in the diagonals of the diagonal matrices $\bD_c$ and $\bD_{cc}$ respectively.
The matrix of weighted scalar products for the rows of $\bZ$ and $\bY$ are then identical.
These $I\times I$ matrices are called the \textit{form matrices}:
\begin{equation}
 \bZ \bD_c \bZ\tr = \bY \bD_{cc} \bY\tr
 \label{FormMatrices}
\end{equation}
thanks to the result $\bP_\textsc{clr}\tr \bD_c \bP_\textsc{clr} = \bP_\textsc{lr}\tr \bD_{cc} \bP_\textsc{lr}$.
This proves that the configuration of the samples is identical.
The distances in (\ref{dist_samples_weighted}) correspond exactly to the form matrices in (\ref{FormMatrices}).

Notice that the equivalence of the PCA of the CLRs and the PCA of the LRs is a particular result thanks to the definition of the CLRs.
In regular PCA the low-dimensional result is not optimal for differences between variables.
For example, \cite{Gabriel:72} gives examples of difference vectors in biplots but makes no statements about their optimality, whereas \cite{Greenacre:03} shows specifically how difference vectors can be optimally displayed.

\section{Discussion}
In compositional data analysis it is all the pairwise logratios that are analysed and interpreted simultaneously.
For multivariate analysis the centered logratio transformation is the most useful and most efficient, since it implies the analysis of all the pairwise logratios.
The matrix of CLRs has $J$ columns but is of rank $J-1$ (it is assumed that the number of rows $I$ is greater than the number of columns $J$, otherwise the rank is $I-1$). 
When a multivariate analysis such as discriminant analysis requires the inversion of a singular covariance matrix, then the generalized inverse needs to be used. 
When there is a response variable and the compositional variables are used as explanatory variables (see Section 2.6), there are important issues of effect-size interpretation, discussed by \cite{Coenders:20}. 

As John Aitchison said, ``Compositional data analysis is simple'' \citep{Aitchison:97}.
The basic concept and initial step is to use the logratio transformation, after which statistical analysis proceeds very much as before.
But care needs to be taken in the interpretation: the analyst has to think in terms of pairwise logratios and realize the implications on the results of the initial normalization of the data to have constant sums.

{\bibliographystyle{plainnat}
\bibliography{Greenacre}}
\end{document}